\pdfoutput=1 
\documentclass[onecolumn,english,american]{sig-alternate}
\usepackage[T1]{fontenc}
\usepackage[utf8]{inputenc}
\usepackage{color}
\usepackage{babel}
\usepackage{array}
\usepackage{textcomp}
\usepackage{amsmath}
\usepackage{amssymb}
\usepackage{graphicx}
\usepackage[unicode=true]{hyperref}

\makeatletter

\providecommand{\tabularnewline}{\\}

\usepackage{hyperref}

\makeatother

\begin{document}
\numberofauthors{3}
\conferenceinfo{IiiX'12}{Nijmegen, The Netherlands.}
\CopyrightYear{2012} 
\crdata{978-1-4503-1282-0/2012/08}  

\title{Ordinary Search Engine Users assessing Difficulty, Effort, and Outcome
for Simple and Complex Search Tasks}

\author{\alignauthor Georg Singer, Ulrich Norbisrath\\
\affaddr{Institute of Computer Science, University of Tartu}\\
\affaddr{Tartu, Estonia}\\
\email{\{FirstName.LastName\}@ut.ee}
\alignauthor Dirk Lewandowski\\
\affaddr{Hamburg University of Applied Sciences}\\
\affaddr{Hamburg, Germany}\\
\email{Dirk.Lewandowski@haw-hamburg.de}}
\maketitle
\begin{abstract}
Search engines are the preferred tools for finding information on
the Web. They are advancing to be the common helpers to answer any
of our search needs. We use them to carry out simple look-up tasks
and also to work on rather time consuming and more complex search
tasks. Yet, we do not know very much about the user performance while
carrying out those tasks -- especially not for ordinary users. The
aim of this study was to get more insight into whether Web users manage
to assess difficulty, time effort, query effort, and task outcome
of search tasks, and if their judging performance relates to task
complexity. Our study was conducted with a systematically selected
sample of 56 people with a wide demographic background. They carried
out a set of 12 search tasks with commercial Web search engines in
a laboratory environment. The results confirm that it is hard for
normal Web users to judge the difficulty and effort to carry out complex
search tasks. The judgments are more reliable for simple tasks than
for complex ones. Task complexity is an indicator for judging performance.
\end{abstract}
\selectlanguage{english}%

\section{Introduction}

People use search engines for all kinds of tasks, from simply looking
up trivia to planning their holiday trips. While looking up dates
and facts usually is an endeavor limited in time and also effort,
more complex tasks usually can take much longer than expected. Carrying
out those tasks with current search engines might also cause much
more effort than expected. This disparity between expected and real
effort for such tasks is partly due to search engines not supporting
those types of tasks explicitly \cite{white_exploratory_2009-1},
but also due to users having little understanding about different
task types. 

Singer et al. \cite{singer_complex_2012} have taken Marchionini's
definition of exploratory search \cite{marchionini_exploratory_2006}
and focus especially on the search concepts discovery, aggregation
and synthesis. Their idea is that these are the most time consuming
activities and they also cause the most search effort when fulfilling
an information need and make a search task complex. They define complex
search tasks as at least requiring one of the elements aggregation
(finding several documents to a known aspect), discovery (detecting
new aspect), and synthesis (synthesizing the found information into
a single document). Complex tasks typically require going through
those steps multiple times. \cite{white_evaluating_2008}. 

According to a key note speech with the title “Search isn't Search”
by Stefan Weitz (Microsoft) given at the SMX Conference 2009 \cite{weitz_search_2009},
only 1 in 4 queries are successful and many queries yield terrible
satisfaction. Many search queries are actually not isolated efforts
towards finding a single fact but instead are part of sessions, close
to 50\% of sessions are longer than 1 week and people are increasingly
using search to make decisions (66\% of search users). 

The above mentioned dissatisfaction of users in terms of unsuccessful
queries might partly be caused by users not being able to judge the
task effort properly and therefore their experience not being in line
with their expectations. The problem is, that little research exists
about a reasonably big sample of ordinary Web search engine users
carrying out complex search tasks and examining their ability to judge
task effort. In this paper, we present a study where a larger number
of ordinary Web search engine users carried out a number of search
tasks (6 simple and 6 complex tasks) in a laboratory environment.
Before each task we asked the users to rate their expectations regarding
task difficulty, task effort and task outcome. Then we had them carry
out the tasks and after each task asked them to do the rating again,
this time them being aware of the real effort.We were especially interested,
if the judging performance varied between simple and complex tasks
and whether good searchers were also better judges. Finally we investigated
whether the judging performance depended on task complexity or rather
simply on the individual person.

The rest of this paper is structured as follows: First, we review
the literature on users estimating complexity of search tasks, followed
by studies on simple, complex (and therefore also covering exploratory)
search. Then, we give some necessary definitions and state our research
questions. After that, we describe our methods, followed by the results.
These are discussed, and in the conclusions section we sum up the
outcomes and limitations of our research and give some directions
for future research.\selectlanguage{american}%

\selectlanguage{english}%

\section{Related Work}

According to Li, ``Tasks are activities people attempt to accomplish
in order to keep their work or life moving on'' (\cite{li_faceted_2008},
p. 1823). Usually tasks have an ultimate goal and information searching
is an activity to find relevant information to achieve that goal \cite{vakkari_task_2003}.
Information searching can be the result of an interest or of a work
task. A work task is a task that appears in the work context. It's
goal is work related \cite{li_faceted_2008}. Work tasks can be the
origin for information-seeking tasks and information search tasks
\cite{bruce_colis_2002,bystrom_conceptual_2005,ingwersen_information_1992,bell_searchers_2004,belkin_ask_1982}.
Li \cite{li_faceted_2008} defines information-seeking tasks as being
related to people's general information needs. Such needs can be satisfied
by searching through multiple sources, including books in libraries,
papers and also digital information systems. Once people only search
with information systems, the information-seeking task becomes a search
task. A family might for example be faced with the task to plan the
holiday trip. Resulting out of this work task, the search task, to
use the Internet and Web search engines to find children-friendly
hotels at a certain destination might arise. 

Task complexity can be either objective or subjective \cite{li_measuring_2011}.
As far as information science is concerned, objective task complexity
is poorly researched. According to Li and Belkin \cite{li_exploration_2010},
task complexity relates to the number of sub-tasks that need to be
carried out. Subjective task complexity reflects how complex the person,
who carries out the task, sees it \cite{li_measuring_2011}. Byström
and Järvelin \cite{bystrom_task_1995} have developed a task categorization
accounting for task complexity from an automatic information processing
task to a genuine decision task. According to Byström and Järvelin
task complexity is mainly defined by users having to deal with ``a
priori determinability of, or uncertainty about, task outcomes, process,
and information requirement'' (\cite{bystrom_task_1995}, p. 194).
They state three types of information needs in tasks: problem information
(specific requirements of the problem dealt with), domain information
(facts, concepts, rules and laws about the domain the problem is located
in) and problem solving information (known methods to tackle this
problem). Their findings show that for automatic information processing
tasks both the level of motivation to carry out the task is high (as
people are quite sure they will be able to solve it) and also only
problem information is needed. In known, genuine decision tasks (which
was the highest level of complexity they investigated in this experiment),
the level of ambition is also high (and higher than expected). They
credit this to the level of education of their study participants.
What clearly distinguished this task from the simple one, was the
level of problem solving information, which was required in terms
of systems to use and experts to ask. 

Bell and Ruthven \cite{bell_searchers_2004} carried out a user study
with 30 people who were asked to work on three groups of search tasks
(tasks organized in three complexity levels) and afterwards rate the
complexity of each task on a 5-point scale. The goal was to test the
above described model by Byström and Järvelin. They observed that
assessment of completion and task complexity were inversely correlated.
The more complex people perceived a task, the less confident they
felt, when they completed that task. In addition they found that a
task is perceived as more complex if the task contains little information
about what information is needed and what amount of information should
be retrieved. Also subjective factors like previous knowledge about
topics related to that task had to be taken into account as influencing
factors for the perception of complexity. 

Gwizdka and Spence \cite{gwizdka_what_2006} conducted a study with
27 undergraduate psychology students in which they where required
to fulfill a look-up task. They wanted to examine the relationship
between searcher's activities and subjective post-task difficulty
and finding predictors for subjective task complexity. They found
that task time, time per click, pages visited, unique pages visited,
revisit ratio and back-button use were good predictors for subjective
task complexity. 

White and Livonen (2002) \cite{white_assessing_2002} conducted a
study with 54 experienced Web searchers and had them rate 16 search
questions regarding complexity. Their results show that users perceive
closed/pre\-dic\-table source questions easy, open/un\-pre\-dictable
source questions difficult. In addition the study participants agreed
that “searchability, clarity, famili\-ari\-ty/curren\-cy, public
knowledge, simplicity, and specificity” were important aspects that
made a task either simple or complex. 

Li et al. \cite{li_measuring_2011} conducted a survey containing
100 university students in China. They observed that objective task
complexity measures were more indicative for task complexity than
subjective ones. The main objective predictors for task complexity
were: number of words in the task description, number of languages
needed to interpret search results and the number of domain areas,
that the task involved. In addition the objective complexity criteria
were more helpful to predict complexity.

In the information science community the two concepts complexity and
difficulty are sometimes used as being identical and sometimes they
are used as being different. Gwizdka \cite{gwizdka_assessing_2010}
has done a question-driven, web-based information search study with
48 participants (students, mean age 27 years) aimed at understanding
the cognitive load when carrying out web search tasks (recording them
and analyzing their respective actions). The study participants were
required to carry out a primary task and in parallel a secondary task
to measure their cognitive load on the primary task. Their results
confirm that subjective task difficulty and objective difficulty are
in line and that study participants tended to underestimate task difficulty.
The author credited this to the high degree of Internet search experience
among the participants and their relatively young age. The study also
shows that subjective difficulty was more strongly related with user
effort than objective difficulty. The author interprets this as the
subjective difficulty more truly showing the searcher's cognitive
effort. Gwizdka's definition of a difficult task has a lot of overlap
with the common definition of a complex task. Yet he really pinpoints
difficulty to cognitive effort, a subject measure that to a large
extend corresponds to subjective complexity.

Vakkari and Huuskonen \cite{vakkari_search_2012} conducted a study
with 41 medical students to investigate how the search effort impacted
search output and task outcome. They found that in case of bad retrieval
results, humans worked harder to achieve desired task outcomes. They
conclude that measures for search process and task outcome need to
be added to classic IR measures.

\section{Definitions}

\subsection*{Objective vs. subjective }

As we have seen, researchers use different concepts to distinguish
simple tasks from more complex ones, either describing them as “complex”
or “difficult”, sometimes interchangeably. To help guide the reader
of this paper, we give some definitions, which we will use throughout
this paper. 

A task is “a usually assigned piece of work often to be finished within
a certain time” (Merriam-Webster). 

A search task is a piece of work concerning the retrieval of information
related to an information need. The search is carried out with search
systems only \cite{li_faceted_2008}.

A search task is complex if it requires at least one of the elements
aggregation, discovery, and synthesis. It typically requires reviewing
many documents and synthesizing them into a desired format. 

A search task is difficult if a lot of cognitive input is needed to
carry out the task.

A search task requires increased effort if the user needs either more
cognitive effort to understand the tasks and formulate queries (time
effort), or more mechanical effort (number of queries, number of pages
visited).\selectlanguage{american}%

\selectlanguage{english}%

\section{Research Questions}

In the study task complexity is the independent variable. Search performance,
searchers' assessments of task difficulty, query effort, time effort
and outcome are dependent variables. 

\label{sec:Research-Questions}To guide our research, we formulated
the following research questions: 
\begin{description}
\item [{RQ1:}] Can users assess difficulty, effort and task outcome for simple
search tasks? 
\item [{RQ2:}] Can users assess difficulty, effort and task outcome for complex
search tasks? 
\item [{RQ3:}] Are there significant performance differences between assessing
simple and complex search tasks? 
\item [{RQ4:}] Does the users' ability to judge if the information they
have found is correct or not depend on task complexity? 
\end{description}
In RQ1 to RQ4 the independent variable is task complexity, the other
variables are dependent variables.
\begin{description}
\item [{RQ5:}] Is there a correlation between the overall search performance
(ranking in the experiment) and the ability to assess difficulty,
time effort, query effort, and task outcome for complex tasks? 
\end{description}
Here we investigate the correlation between search performance and
judging performance.
\begin{description}
\item [{RQ6:}] Does the judging performance depend on task complexity or
simply the individual user?
\end{description}
This question investigates the association between judging performance,
task complexity and the individual user.\selectlanguage{american}%

\selectlanguage{english}%

\section{Research Method}

The results presented in this paper are based on a body of data gathered
in the course of a larger experiment in August 2011. One additional
article using distinct parts of the data and describing different
aspects has been published \cite{singer_relationship_2012}, one article
is currently in press \cite{singer_impact_2012}, and one article
has been submitted for review \cite{singer_experiment_2012}. The
following description of the research design is based on Singer et
al. \cite{singer_experiment_2012}. The experiment was conducted in
August 2011 in Hamburg, Germany. Participants were invited to the
university, where they were given some search tasks (see below). The
study was carried out in one of the university's computer labs, where
each participant had her own computer and was instructed to work on
the search tasks independently. Participants were not observed directly,
but their browser interactions were recorded using the Search-Logger
plug-in. While tasks were presented in a certain order to the participants,
they were allowed to choose the order of the tasks according to their
wishes, and it was also possible to interrupt a task, work on another
one, and later return.

We recruited a sample of 60 volunteers, using a demographic structure
model. The aim was to go beyond the usual user samples consisting
mainly of students, often experienced searchers from information science
or computer science, and also, to increase the sample size. As a user
sample of the intended size could not be representative, we wanted
at least to make sure that adults from various age ranges, and also
men and women alike, were considered. For details on the sample, see
Table~\ref{tab:Demography-of-user-sample}. The effective number
of study participants providing data to our study was reduced to 56,
as the data of 4 (2 females, 2 males) out of the 60 users was corrupt
and could therefore not be used. 
\begin{table}
\begin{centering}
\begin{tabular}{|c|c|c|c|}
\hline 
\textbf{Basic Data} & \multicolumn{2}{c|}{\textbf{Gender}} & \tabularnewline
\hline 
\hline 
\textbf{Age Span} & \textbf{Female} & \textbf{Male} & \textbf{Total}\tabularnewline
\hline 
\hline 
18-24 & 5 & 4 & 9\tabularnewline
\hline 
25-34 & 9 & 7 & 16\tabularnewline
\hline 
35-44 & 7 & 8 & 15\tabularnewline
\hline 
45-54 & 8 & 8 & 16\tabularnewline
\hline 
55-59 & 3 & 1 & 4\tabularnewline
\hline 
\textbf{Total} & \textbf{32} & \textbf{28} & \textbf{60}\tabularnewline
\hline 
\end{tabular}
\par\end{centering}

\caption{Demography of user sample\label{tab:Demography-of-user-sample}}
\end{table}

The search experiment consisted of 12 search tasks. As our experiment
was conducted in Germany, the language of the tasks was German. A
prerequisite for all tasks was that a correct answer had to be available
somewhere in public websites in German as of August 2011. The study
participants had 3 hours to complete the experiment. 

Simple tasks were characterized by asking the users to find simple
facts. The needed information was contained in one document (web site)
and could be retrieved with one single query. Complex tasks on the
other hand were formulated in a way that the user had enough context
to comprehend the task situation but were still characterized by uncertainty
and ambiguity \cite{kules_designing_2009}. There was no single right
answer retrievable and the required information was spread over various
documents (web sites). Fulfilling the task typically required issuing
multiple queries, aggregating information from various sources and
synthesizing the information into a single solution document \cite{singer_complex_2012}.
The tasks were as follows (in a mixed order, (S) marks simple, and
(C) complex tasks):
\begin{enumerate}
\item \textbf{(S)} When was the composer of the piece “The Magic Flute”
born?
\item \textbf{(S)} How high is the state debt of Italy in comparison to
their gross domestic product (GDP) in June 2011 in \%?
\item \textbf{(S)} How many opera pieces did Verdi composed?
\item \textbf{(S)} When and by whom was penicillin discovered?
\item \textbf{(S)} Joseph Pulitzer (1847-1911) was a well-known journalist
and publisher from the U.S. The Pulitzer Prize carries on his name.
In which European country was Pulitzer born?
\item \textbf{(S)} How many Euros do you get if you exchange 10.000 units
of the currency of Lithuania?
\item \textbf{(S)} How hot can it be on average in July in Aachen/Germany?
\item \textbf{(C)} What are the most important five points to consider if
you want to plan a budget wedding?
\item \textbf{(C)} You were offered the job to run the local Goethe Institute
(responsible for German language and cultural education) abroad. The
chance is high that you will be sent to Astana (Kazakhstan). Please
collect facts and information (about half a page) about the political
situation in Kazakhstan and the living quality.
\item \textbf{(C)} What is the name of the creature on the following picture
and who is the author? Hint: this Austrian writer is also well known
in Germany.
\item \textbf{(C)} Are there differences regarding the distribution of religious
affiliations between Austria, Germany, and Switzerland? Which ones?
\item \textbf{(C)} There are five countries whose names are also carried
by chemical elements. France has two (31. Ga – Gallium and 87. Fr
– Frantium), Germany has one (32. Ge – Germanium), Russia has one
(44. Ru – Rutentium) and Poland has one (84. Po – Polonium). Please
name the fifth country.
\end{enumerate}
We set the sequence of tasks up in way, so that users could alternatively
solve simple and complex ones. The aim was to keep the participants
interested, and to not discourage participants through a sequence
of complex search tasks which they might be unable to solve.

We implemented users' judgments as binary responses to questionnaire
items. We added questionnaires before starting and after finishing
each search task. Prior to each task we used the following statements,
that users could rate with yes or no: 1) This task is easy 2) It will
take me less than 5 minutes to complete the task 3) I will need fewer
than 5 queries to complete the task 4) I will find the correct information.
After the participants had completed that task we asked them to rate
the following statements with yes or no: 1) The task was easy 2) I
took me less than 5 minutes to complete the task 3) It needed fewer
than 5 queries to complete the task 4) I have found the correct information. 

In terms of effort, we understood effort as comprising the cognitive
effort to understand the task and formulate queries (time) and also
the mechanical effort to carry out the task (number of queries). This
is different than in Jacek Gwizdka~\cite{gwizdka_assessing_2010}.
In his study, he measures the cognitive load or mental effort. For
our study, measuring cognitive effort directly was not feasible, therefore
we measured the indicators time effort and query effort.

We compared users’ subjective values for the question whether they
thought in advance they would find the correct information (yes or
no) with the objectively graded outcome that they submitted. The objective
result is a manual review of all the answers given by the participants
of our study. The solutions the study participants provided were bench-marked
against this optimal solution developed by the researchers on a scale,
correct, partly correct, wrong, no solution submitted. In case of
simple tasks, each task had exactly one solution. If the solution
was correct and complete, the task was graded correct. If it was correct
and not complete (like only mentioning who invented Penicillin but
not when in above mentioned Task 4), it was graded ``partly correct''.
If the solution was wrong, it was graded ``incorrect''. In the case
of complex tasks, it was less trivial. As the tasks were quite open,
there was no single right or wrong solution possible. If the solution
provided by the users covered all aspects that the optimal solution
also contained it was marked ``correct''. If the solution covered
fewer aspects it was graded ``partly correct''. If the solution
did not cover any aspects the solution was ``incorrect''. For both,
simple and complex tasks, if no solution was submitted, the task was
graded ``unanswered''.

To understand the relation between search performance (ranking of
the user in the experiment) and the ability to estimate task difficulty,
task effort, and task outcome, we ordered the users according to their
ranking in the experiment. We ranked the users first by the number
of correct answers and then, in cases of users with the same number
of correct answers, by ``partly correct'' answers. 

We ran paired-sample t-tests (assuming unequal variances) to analyze
the statistical significance of our results for RQ3- RQ5.\selectlanguage{american}%

\selectlanguage{english}%

\section{Results}

In this section we present the results of our study, which are also
use to answer the research questions stated in Section~\ref{sec:Research-Questions}.
We first show, how ordinary Web search engine users manage to judge
the difficulty, effort and task outcome for simple tasks. Next we
present the results of the same analysis but this time applied to
complex tasks. Then we highlight the differences in judging performance
between simple and complex tasks. We show, how task complexity impacts
the ability of the users to judge whether the information they have
found is correct or not. Next we examine, if better searchers are
better at judging than worse searchers. Finally we investigate whether
the judging performance depends on the task complexity or the individual
user only.

It is important to point out that the numbers that we are presenting
here, are subjective difficulty, subjective effort, and subjective
ability to find the correct information. Subjective here means the
individual judgment of the searcher in answering a question of our
pre-task and post-task questionnaires. If a task was subjectively
difficult, this does not necessarily mean that this would also be
the case on an objective level as outlined in the related work section~\cite{bystrom_task_1995}.

\subsection*{RQ1: Can users assess difficulty, effort and task outcome for simple
search tasks? }

Table~\ref{tab:Users-estimating-simple} outlines our findings regarding
users and their ability to estimate difficulty, effort (in terms of
time used and queries entered) and being able to find the correct
result for simple search tasks. ``\textbf{\# of tasks}'' represents
the number of simple tasks that have been processed by the study participants.
The total number of tasks (correct plus incorrect ones) should have
been 56{*}6=336 (56 valid users x 6 tasks). It is slightly lower due
to invalid or not given answers or not fulfilled tasks. \textbf{\%}
shows the percentage of the number of judged tasks to the total valid
answers for tasks. We graded an answer as correct when the users'
self-judged values were the same in the pre-task questionnaire and
the post-task questionnaire. For example if they judged a task to
be difficult in the pre-task questionnaire and after carrying out
the task stated again that it was a difficult task, the judgment was
graded as correct.

For all parameters (difficulty, time effort, query effort, and result
finding ability) approximately 90\% of the users managed to match
estimated and experienced values for simple tasks. However, in our
study the users had slightly more trouble estimating the time effort
needed than the query effort (in terms of estimating going over a
threshold of numbers of queries).

\begin{table}
\begin{centering}
\begin{tabular}{|c|c|c|}
\cline{2-3} 
\multicolumn{1}{c|}{} & \textbf{\# of tasks} & \textbf{\%}\tabularnewline
\hline 
\hline 
\textbf{difficulty} &  & \tabularnewline
\hline 
\hline 
incorrect & 29 & 9.8\tabularnewline
\hline 
correct & 266 & 90.2\tabularnewline
\hline 
\multicolumn{1}{c}{} & \multicolumn{1}{c}{} & \multicolumn{1}{c}{}\tabularnewline
\hline 
\textbf{time effort} &  & \tabularnewline
\hline 
\hline 
incorrect & 27 & 9.1\tabularnewline
\hline 
correct & 268 & 90.8\tabularnewline
\hline 
\multicolumn{1}{c}{} & \multicolumn{1}{c}{} & \multicolumn{1}{c}{}\tabularnewline
\hline 
\textbf{query effort} &  & \tabularnewline
\hline 
\hline 
incorrect & 38 & 12.9\tabularnewline
\hline 
correct & 257 & 87.1\tabularnewline
\hline 
\multicolumn{1}{c}{} & \multicolumn{1}{c}{} & \multicolumn{1}{c}{}\tabularnewline
\hline 
\textbf{ability to find right result} &  & \tabularnewline
\hline 
\hline 
incorrect & 16 & 5.4\tabularnewline
\hline 
correct & 279 & 94.6\tabularnewline
\hline 
\end{tabular}
\par\end{centering}

\caption{\label{tab:Users-estimating-simple}Users judging simple search tasks}
\end{table}

\subsection*{RQ2: Can users assess difficulty, effort and task outcome for complex
search tasks? }

Table \ref{tab:Users-estimating-complex} outlines our findings regarding
users and their ability to estimate difficulty, efforts, and being
able to find the correct result for complex search tasks. For all
parameters to estimate, about 70\% of all tasks were judged correctly
with slightly worse estimations for query effort and result-finding
skills.

\begin{table}
\begin{centering}
\begin{tabular}{|c|c|c|}
\cline{2-3} 
\multicolumn{1}{c|}{} & \textbf{\# of tasks} & \textbf{\%}\tabularnewline
\hline 
\hline 
\textbf{difficulty} &  & \tabularnewline
\hline 
\hline 
incorrect & 95 & 33.2\tabularnewline
\hline 
correct & 191 & 66.8\tabularnewline
\hline 
\multicolumn{1}{c}{} & \multicolumn{1}{c}{} & \multicolumn{1}{c}{}\tabularnewline
\hline 
\textbf{time effort} &  & \tabularnewline
\hline 
\hline 
incorrect & 99 & 34.6\tabularnewline
\hline 
correct & 187 & 65.3\tabularnewline
\hline 
\multicolumn{1}{c}{} & \multicolumn{1}{c}{} & \multicolumn{1}{c}{}\tabularnewline
\hline 
\textbf{query effort} &  & \tabularnewline
\hline 
\hline 
incorrect & 91 & 31.8\tabularnewline
\hline 
correct & 195 & 68.2\tabularnewline
\hline 
\multicolumn{1}{c}{} & \multicolumn{1}{c}{} & \multicolumn{1}{c}{}\tabularnewline
\hline 
\textbf{ability to find right result} &  & \tabularnewline
\hline 
\hline 
incorrect & 78 & 27.2\tabularnewline
\hline 
correct & 208 & 72.8\tabularnewline
\hline 
\end{tabular}
\par\end{centering}

\caption{Users judging complex search tasks\label{tab:Users-estimating-complex}}
\end{table}

\subsection*{RQ3: Are there significant differences between assessing simple and
complex search tasks?}

To analyze if there exist significant differences between how users
estimate certain parameters (difficulty, time effort, query effort,
search success), we have compared the differences between users’ pre-task
estimate and post-task experience based values for 295 simple tasks
and 286 complex tasks as outlined in Table \ref{tab:simple-versus-complex}
(100\% means a 100\% probability to judge the right difficulty; pre-task
estimate and post task evaluation are totally in line). We used paired
sample t-tests to compare the results from simple and complex tasks.
We paired average difficulty for simple tasks and average difficulty
for complex tasks. We followed the same procedure for time effort,
query effort and task outcome.

In the case of simple tasks, users are significantly better at estimating
all four parameters: difficulty, time effort, query effort and search
success, i.e. the difference between their pre-task estimate and their
post-task experience based value for a certain parameter is significantly
smaller and they show a higher probability to correctly judge the
parameter. Table~\ref{tab:parameter-per-task} shows the judging
performance for all users for each specific task. Also here the division
in terms of performance between simple \textbf{(S)} and complex \textbf{(C)}
tasks is clearly visible. 

\begin{table}
\begin{centering}
\begin{tabular*}{1\columnwidth}{@{\extracolsep{\fill}}|>{\centering}p{0.15\columnwidth}|>{\centering}p{0.15\columnwidth}|>{\centering}p{0.15\columnwidth}|>{\centering}p{0.15\columnwidth}|>{\centering}p{0.15\columnwidth}|}
\hline 
 & difficulty (\%) & time effort (\%) & query effort (\%) & task outcome (\%)\tabularnewline
\hline 
\hline 
Simple tasks (n=295)  & 90\textpm{}2 & 91\textpm{}2 & 87\textpm{}2 & 95\textpm{}1\tabularnewline
\hline 
Complex tasks (n=286)  & 67\textpm{}3 & 65\textpm{}3 & 68\textpm{}3 & 73\textpm{}3\tabularnewline
\hline 
p-value  & <0.001 (s) & <0.001 (s) & <0.001 (s) & <0.001 (s)\tabularnewline
\hline 
\end{tabular*}
\par\end{centering}

\begin{raggedleft}
s...significant
\par\end{raggedleft}

\caption{Correct judged tasks per dependent variable (mean values over tasks)
\label{tab:simple-versus-complex}}
\end{table}

\begin{table}
\begin{centering}
\begin{tabular}{|>{\centering}p{0.25\columnwidth}|>{\centering}p{0.13\columnwidth}|>{\centering}p{0.13\columnwidth}|>{\centering}p{0.13\columnwidth}|>{\centering}p{0.13\columnwidth}|}
\hline 
Task & difficulty (\%) & time effort (\%) & query effort (\%) & task outcome (\%)\tabularnewline
\hline 
\hline 
1 \textbf{(S)} (n=51) & 92\textpm{}4 & 90\textpm{}4 & 88\textpm{}5 & 98\textpm{}2\tabularnewline
\hline 
2 \textbf{(S)} (n=48) & 81\textpm{}6 & 85\textpm{}5 & 73\textpm{}6 & 88\textpm{}5\tabularnewline
\hline 
5 \textbf{(S)} (n=47) & 89\textpm{}5 & 91\textpm{}4 & 87\textpm{}5 & 96\textpm{}3\tabularnewline
\hline 
8 \textbf{(S)} (n=51) & 98\textpm{}2 & 100\textpm{}0 & 98\textpm{}3 & 96\textpm{}3\tabularnewline
\hline 
9 \textbf{(S)} (n=49) & 84\textpm{}5 & 82\textpm{}6 & 82\textpm{}6 & 94\textpm{}3\tabularnewline
\hline 
12 \textbf{(S)} (n=49) & 96\textpm{}3 & 96\textpm{}3 & 94\textpm{}3 & 96\textpm{}3\tabularnewline
\hline 
3 \textbf{(C)} (n=47) & 62\textpm{}7 & 51\textpm{}7 & 60\textpm{}7 & 85\textpm{}5\tabularnewline
\hline 
4 \textbf{(C)} (n=48) & 67\textpm{}7 & 69\textpm{}7 & 71\textpm{}7 & 77\textpm{}6\tabularnewline
\hline 
6 \textbf{(C)} (n=48) & 60\textpm{}7 & 81\textpm{}6 & 73\textpm{}6 & 81\textpm{}6\tabularnewline
\hline 
7 \textbf{(C)} (n=46) & 72\textpm{}7 & 67\textpm{}7 & 72\textpm{}7 & 52\textpm{}7\tabularnewline
\hline 
10 \textbf{(C)} (n=49) & 65\textpm{}7 & 67\textpm{}7 & 65\textpm{}7 & 61\textpm{}7\tabularnewline
\hline 
11 \textbf{(C)} (n=47) & 74\textpm{}6 & 55\textpm{}7 & 68\textpm{}7 & 79\textpm{}6\tabularnewline
\hline 
\end{tabular}
\par\end{centering}

\caption{Fraction of users correctly judging task parameters per task \label{tab:parameter-per-task}}
\end{table}

\subsection*{RQ4: Does the users' ability to judge if the information they have
found is correct or not depend on task complexity? }

We compared users’ subjective values for the question whether they
think in advance they would find the correct information (yes or no)
with the objectively graded outcome that they submitted (by building
the difference of the submitted values). The objective result is a
manual review of all the answers given by the participants of our
study. In this evaluation we skipped the tasks where a user had not
delivered any result as we did not know what was the reason for not
delivering (could be not being able, found wrong results and did not
want to submit, or simply forgot to submit). We also analyzed, whether
the users were able to judge the correctness of their found results
after having finished the search task. 

These evaluations gives us an estimate of how well users can judge
that a result they found on the Internet is actually correct and how
well they can judge in advance, if they will be able to find the correct
result.

Table~\ref{tab:Absolute-differences-between} shows that the ability
to predict, whether it is possible to find the correct information
is significantly higher for simple search tasks than for complex search
tasks, 87\% versus 52\%. We used paired sample t-tests to compare
the results from simple and complex tasks. We paired average correctness
(difference user estimated vs. objectively graded outcome) for simple
tasks and average correctness for complex tasks.

Table~\ref{tab:Absolute-differences-between-1} depicts that the
ability to judge whether a found information is correct or not is
significantly higher in case of simple tasks than it is in case of
complex tasks. Here the difference is also significant, 88\% versus
60\%. 

The difference of observations between simple and complex tasks is
due to the fact that not all users always correctly entered their
rating into our system and therefore those tasks had to be omitted.

\begin{table}
\begin{centering}
\begin{tabular}{|>{\centering}p{0.25\columnwidth}|>{\centering}p{0.4\columnwidth}|}
\hline 
Task type & Correctly estimated tasks (\%)\tabularnewline
\hline 
\hline 
simple (n=259) & 87\textpm{}2\tabularnewline
\hline 
complex (n=233) & 52\textpm{}3\tabularnewline
\hline 
p-value & <0.001 (s)\tabularnewline
\hline 
\end{tabular}
\par\end{centering}

\caption{Correct judgments of expected (pre-task questionnaire) search results
compared to correctness of manually evaluated search results (mean
values over tasks)\label{tab:Absolute-differences-between}}
\end{table}

\begin{table}
\begin{centering}
\begin{tabular}{|>{\centering}p{0.25\columnwidth}|>{\centering}p{0.4\columnwidth}|}
\hline 
Task type & Correctly estimated tasks (\%)\tabularnewline
\hline 
\hline 
simple (n=259) & 88\textpm{}2\tabularnewline
\hline 
complex (n=230) & 60\textpm{}3\tabularnewline
\hline 
p-value & <0.001 (s)\tabularnewline
\hline 
\end{tabular}
\par\end{centering}

\caption{Correct assessments of self-judged (post-task questionnaire) search
results compared to correctness of manually evaluated search results
(mean values over tasks)\label{tab:Absolute-differences-between-1}}
\end{table}

\subsection*{RQ5: Is there a correlation between the overall search performance
(ranking in the experiment) and the ability to assess difficulty,
time effort, query effort, and task outcome for complex tasks? }

As also mentioned in the methods section, to understand the relation
between search performance (ranking of the user in the experiment)
and the ability to estimate task difficulty, task effort, and task
outcome, we ordered the users according to their ranking in the experiment.
We ranked the users first by the number of correct answers given and
then, in cases of users with the same number of correct answers, by
answers with right elements (simple and complex tasks). 

Then we compared the complex tasks of the first quartile of ranked
users (n=67 tasks) with the complex tasks of the fourth quartile of
ranked users (n=59 tasks) as outlined in Table \ref{tab:Differences-between-expected and experienced}.
The number of tasks is different due to the fact that not all users
always correctly entered their estimates into the system and therefore
those tasks had to be omitted.

\begin{table}
\begin{centering}
\begin{tabular*}{1\columnwidth}{@{\extracolsep{\fill}}|>{\centering}p{0.15\columnwidth}|>{\centering}p{0.15\columnwidth}|>{\centering}p{0.15\columnwidth}|>{\centering}p{0.15\columnwidth}|>{\centering}p{0.15\columnwidth}|}
\hline 
 &  Avg. difficulty in \% & Avg.

time effort in \% & Avg. query effort in \% & Avg. task outcome in \%\tabularnewline
\hline 
\hline 
1. quartile (n=67)  & 67\textpm{}6 & 64\textpm{}6 & 67\textpm{}6 & 85\textpm{}4\tabularnewline
\hline 
4. quartile (n=59)  & 73\textpm{}6 & 75\textpm{}6 & 73\textpm{}6 & 64\textpm{}6\tabularnewline
\hline 
p-value  & n.s. & n.s. & n.s. & <0.05 (s)\tabularnewline
\hline 
\end{tabular*}
\par\end{centering}

\caption{Correct estimations of best and worst quartile for expected and experienced
task parameters\label{tab:Differences-between-expected and experienced}}
\end{table}

The results show that good searchers are not significantly better
at judging difficulty and effort for complex search tasks. However,
they are significantly better at judging the task outcome.

\subsection*{RQ6: Does the judging performance depend on task complexity or simply
on the individual user?}

In this subsection we examine whether the judging performance depends
on task complexity or simply on the individual ability to make those
judgments. It could for example be that some users are very good at
judging simple as well as complex tasks while others perform badly
for both task groups. In an analysis that takes the average over all
simple tasks and compares them with the average over all complex tasks
(independent of the user e.g. as done for research question 3) this
fact would not show up. 

Table~\ref{tab:Users-judging-the-difficulty} shows the results for
users judging the task difficulty for simple and complex tasks at
the same time. 3 out of 53 users (6\%) were able to judge the difficulty
totally right for simple and complex tasks. 30 out of 53 users (56\%)
managed to judge the difficulty for simple tasks right and at the
same time were not totally correct for the complex tasks. Three users
(6\%) were able to correctly judge all complex tasks while at the
same time wrongly judging a number of simple tasks.

\begin{table}
\begin{tabular}{|>{\raggedright}p{0.03\columnwidth}|>{\raggedright}p{0.06\columnwidth}|>{\raggedright}p{0.08\columnwidth}||>{\raggedright}p{0.03\columnwidth}|>{\raggedright}p{0.06\columnwidth}|>{\raggedright}p{0.08\columnwidth}||>{\raggedright}p{0.03\columnwidth}|>{\raggedright}p{0.06\columnwidth}|>{\raggedright}p{0.08\columnwidth}|}
\hline 
{\tiny user} & {\tiny simple } & {\tiny complex} & {\tiny user} & {\tiny simple} & {\tiny complex} & {\tiny user} & {\tiny simple} & {\tiny complex}\tabularnewline
\hline 
\hline 
{\tiny 44} & {\tiny 100\%} & {\tiny 100\%} & {\tiny 65} & {\tiny 100\%} & {\tiny 67\%} & {\tiny 87} & {\tiny 83\%} & {\tiny 83\%}\tabularnewline
\hline 
{\tiny 58} & {\tiny 100\%} & {\tiny 100\%} & {\tiny 71} & {\tiny 100\%} & {\tiny 67\%} & {\tiny 55} & {\tiny 83\%} & {\tiny 75\%}\tabularnewline
\hline 
{\tiny 92} & {\tiny 100\%} & {\tiny 100\%} & {\tiny 78} & {\tiny 100\%} & {\tiny 67\%} & {\tiny 82} & {\tiny 83\%} & {\tiny 67\%}\tabularnewline
\hline 
{\tiny 28} & {\tiny 100\%} & {\tiny 83\%} & {\tiny 90} & {\tiny 100\%} & {\tiny 67\%} & {\tiny 39} & {\tiny 83\%} & {\tiny 50\%}\tabularnewline
\hline 
{\tiny 42} & {\tiny 100\%} & {\tiny 83\%} & {\tiny 48} & {\tiny 100\%} & {\tiny 60\%} & {\tiny 70} & {\tiny 83\%} & {\tiny 33\%}\tabularnewline
\hline 
{\tiny 46} & {\tiny 100\%} & {\tiny 83\%} & {\tiny 37} & {\tiny 100\%} & {\tiny 50\%} & {\tiny 77} & {\tiny 83\%} & {\tiny 33\%}\tabularnewline
\hline 
{\tiny 61} & {\tiny 100\%} & {\tiny 83\%} & {\tiny 45} & {\tiny 100\%} & {\tiny 50\%} & {\tiny 81} & {\tiny 80\%} & {\tiny 100\%}\tabularnewline
\hline 
{\tiny 63} & {\tiny 100\%} & {\tiny 83\%} & {\tiny 57} & {\tiny 100\%} & {\tiny 50\%} & {\tiny 59} & {\tiny 80\%} & {\tiny 80\%}\tabularnewline
\hline 
{\tiny 66} & {\tiny 100\%} & {\tiny 83\%} & {\tiny 67} & {\tiny 100\%} & {\tiny 50\%} & {\tiny 72} & {\tiny 80\%} & {\tiny 50\%}\tabularnewline
\hline 
{\tiny 89} & {\tiny 100\%} & {\tiny 83\%} & {\tiny 68} & {\tiny 100\%} & {\tiny 50\%} & {\tiny 75} & {\tiny 67\%} & {\tiny 100\%}\tabularnewline
\hline 
{\tiny 34} & {\tiny 100\%} & {\tiny 80\%} & {\tiny 73} & {\tiny 100\%} & {\tiny 50\%} & {\tiny 47} & {\tiny 67\%} & {\tiny 83\%}\tabularnewline
\hline 
{\tiny 43} & {\tiny 100\%} & {\tiny 80\%} & {\tiny 88} & {\tiny 100\%} & {\tiny 50\%} & {\tiny 64} & {\tiny 67\%} & {\tiny 50\%}\tabularnewline
\hline 
{\tiny 54} & {\tiny 100\%} & {\tiny 80\%} & {\tiny 32} & {\tiny 100\%} & {\tiny 40\%} & {\tiny 79} & {\tiny 67\%} & {\tiny 50\%}\tabularnewline
\hline 
{\tiny 93} & {\tiny 100\%} & {\tiny 80\%} & {\tiny 84} & {\tiny 100\%} & {\tiny 40\%} & {\tiny 83} & {\tiny 67\%} & {\tiny 50\%}\tabularnewline
\hline 
{\tiny 24} & {\tiny 100\%} & {\tiny 67\%} & {\tiny 52} & {\tiny 100\%} & {\tiny 20\%} & {\tiny 33} & {\tiny 60\%} & {\tiny 60\%}\tabularnewline
\hline 
{\tiny 38} & {\tiny 100\%} & {\tiny 67\%} & {\tiny 76} & {\tiny 83\%} & {\tiny 40\%} & {\tiny 74} & {\tiny 50\%} & {\tiny 60\%}\tabularnewline
\hline 
{\tiny 51} & {\tiny 100\%} & {\tiny 67\%} & {\tiny 41} & {\tiny 83\%} & {\tiny 83\%} & {\tiny 69} & {\tiny 40\%} & {\tiny 100\%}\tabularnewline
\hline 
{\tiny 62} & {\tiny 100\%} & {\tiny 67\%} & {\tiny 60} & {\tiny 83\%} & {\tiny 83\%} &  &  & \tabularnewline
\hline 
\end{tabular}

\caption{\label{tab:Users-judging-the-difficulty}Users judging the task difficulty }
 
\end{table}

\begin{figure}
\includegraphics[bb=75bp 68bp 650bp 535bp,clip,width=1\columnwidth]{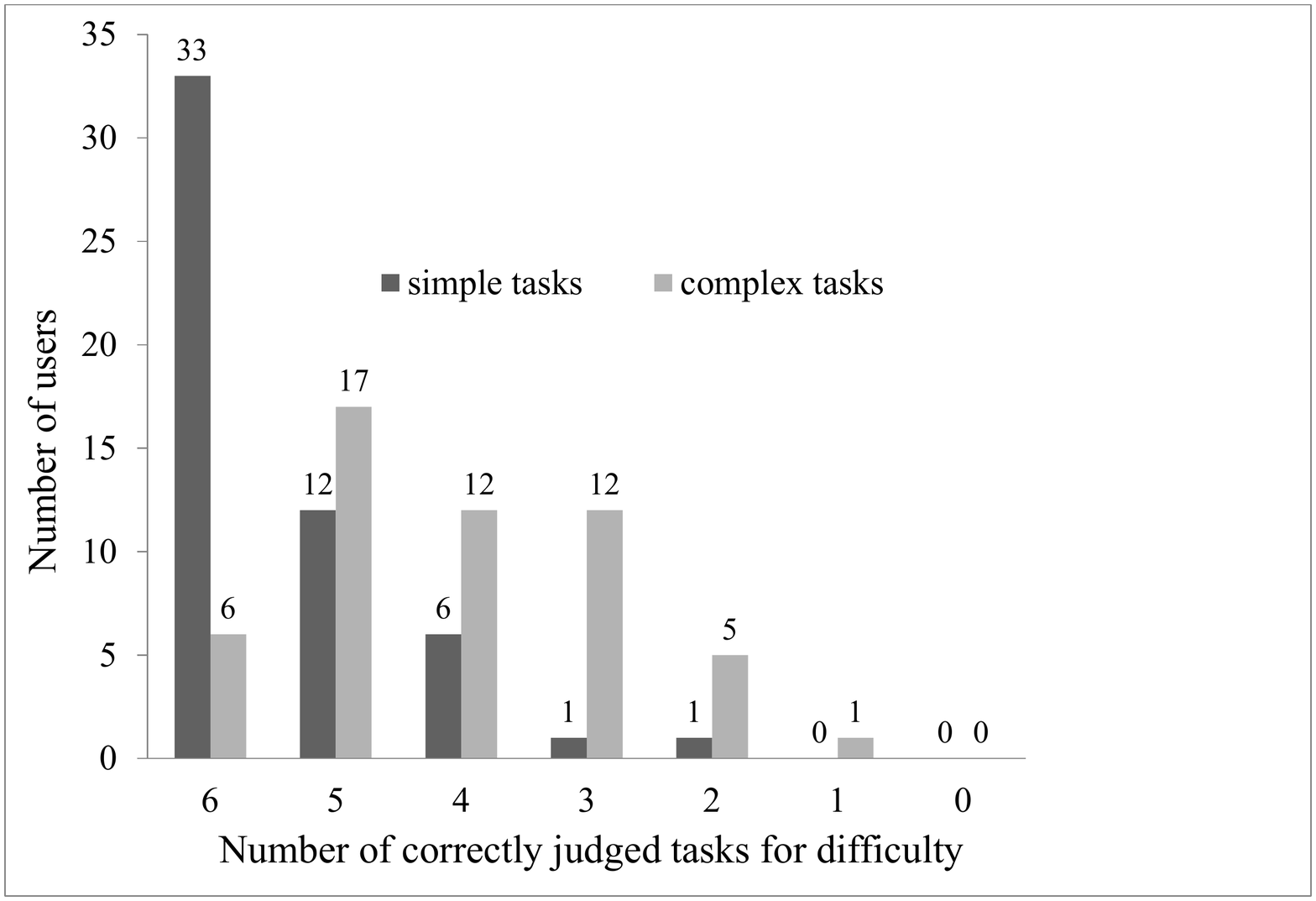}

\caption{\label{fig:Users-judging-the-difficulty}Users judging the task difficulty}
\end{figure}

Figure~\ref{fig:Users-judging-the-difficulty} illustrates a histogram
users versus correctly judged task difficulty. The x-axis shows the
number of correctly carried out tasks from 6 (all) to 0 (none). The
y-axis shows number of users (black bar simple tasks, grey bar complex).
From this figure it is also evident that users are far better at judging
the difficulty for simple tasks. 33 users of 53 (62\%) have managed
to correctly judge the difficulty of all simple tasks, but only 6
out 53 (11\%) have managed to correctly judge the difficulty of all
complex tasks.

Table~\ref{tab:Users-judging-the-time-effort} shows the results
for users judging the time effort for simple and complex tasks. 4
out of 53 users (8\%) were able to judge the time effort right for
simple and complex tasks. 31 out of 53 users (58\%) managed to judge
the time effort for simple tasks right and at the same time were not
totally correct for the complex tasks. One user was able to correctly
judge all complex tasks while wrongly judging a number of simple tasks.

\begin{figure}
\includegraphics[bb=90bp 68bp 650bp 535bp,clip,width=1\columnwidth]{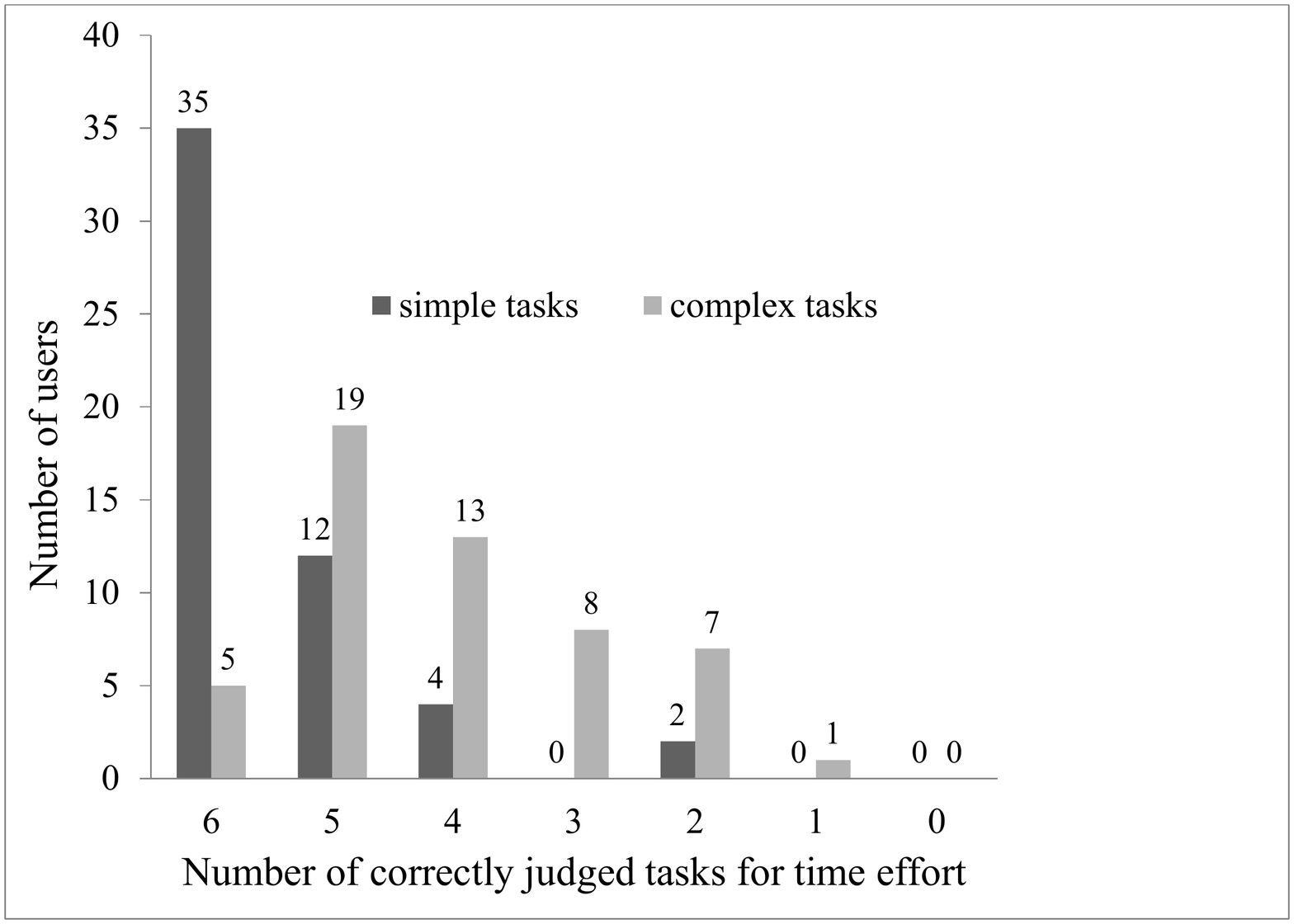}

\caption{\label{fig:Users-judging-the-time-effort}Users judging the time effort}
\end{figure}

\begin{table}
\begin{tabular}{|>{\raggedright}p{0.03\columnwidth}|>{\raggedright}p{0.06\columnwidth}|>{\raggedright}p{0.08\columnwidth}||>{\raggedright}p{0.03\columnwidth}|>{\raggedright}p{0.06\columnwidth}|>{\raggedright}p{0.08\columnwidth}||>{\raggedright}p{0.03\columnwidth}|>{\raggedright}p{0.06\columnwidth}|>{\raggedright}p{0.08\columnwidth}|}
\hline 
{\tiny user} & {\tiny simple } & {\tiny complex} & {\tiny user} & {\tiny simple} & {\tiny complex} & {\tiny user} & {\tiny simple} & {\tiny complex}\tabularnewline
\hline 
\hline 
{\tiny 61} & {\tiny 100\%} & {\tiny 100\%} & {\tiny 63} & {\tiny 100\%} & {\tiny 67\%} & {\tiny 59} & {\tiny 83\%} & {\tiny 83\%}\tabularnewline
\hline 
{\tiny 69} & {\tiny 100\%} & {\tiny 100\%} & {\tiny 82} & {\tiny 100\%} & {\tiny 67\%} & {\tiny 72} & {\tiny 83\%} & {\tiny 83\%}\tabularnewline
\hline 
{\tiny 71} & {\tiny 100\%} & {\tiny 100\%} & {\tiny 92} & {\tiny 100\%} & {\tiny 67\%} & {\tiny 24} & {\tiny 83\%} & {\tiny 80\%}\tabularnewline
\hline 
{\tiny 83} & {\tiny 100\%} & {\tiny 100\%} & {\tiny 33} & {\tiny 100\%} & {\tiny 60\%} & {\tiny 75} & {\tiny 83\%} & {\tiny 67\%}\tabularnewline
\hline 
{\tiny 37} & {\tiny 100\%} & {\tiny 83\%} & {\tiny 38} & {\tiny 100\%} & {\tiny 60\%} & {\tiny 48} & {\tiny 83\%} & {\tiny 50\%}\tabularnewline
\hline 
{\tiny 42} & {\tiny 100\%} & {\tiny 83\%} & {\tiny 64} & {\tiny 100\%} & {\tiny 60\%} & {\tiny 41} & {\tiny 83\%} & {\tiny 40\%}\tabularnewline
\hline 
{\tiny 44} & {\tiny 100\%} & {\tiny 83\%} & {\tiny 66} & {\tiny 100\%} & {\tiny 60\%} & {\tiny 47} & {\tiny 83\%} & {\tiny 33\%}\tabularnewline
\hline 
{\tiny 57} & {\tiny 100\%} & {\tiny 83\%} & {\tiny 28} & {\tiny 100\%} & {\tiny 50\%} & {\tiny 78} & {\tiny 83\%} & {\tiny 25\%}\tabularnewline
\hline 
{\tiny 65} & {\tiny 100\%} & {\tiny 83\%} & {\tiny 34} & {\tiny 100\%} & {\tiny 50\%} & {\tiny 46} & {\tiny 80\%} & {\tiny 80\%}\tabularnewline
\hline 
{\tiny 68} & {\tiny 100\%} & {\tiny 83\%} & {\tiny 51} & {\tiny 100\%} & {\tiny 50\%} & {\tiny 55} & {\tiny 80\%} & {\tiny 75\%}\tabularnewline
\hline 
{\tiny 79} & {\tiny 100\%} & {\tiny 83\%} & {\tiny 73} & {\tiny 100\%} & {\tiny 50\%} & {\tiny 54} & {\tiny 75\%} & {\tiny 60\%}\tabularnewline
\hline 
{\tiny 89} & {\tiny 100\%} & {\tiny 83\%} & {\tiny 76} & {\tiny 100\%} & {\tiny 50\%} & {\tiny 67} & {\tiny 67\%} & {\tiny 80\%}\tabularnewline
\hline 
{\tiny 39} & {\tiny 100\%} & {\tiny 80\%} & {\tiny 87} & {\tiny 100\%} & {\tiny 50\%} & {\tiny 58} & {\tiny 67\%} & {\tiny 40\%}\tabularnewline
\hline 
{\tiny 81} & {\tiny 100\%} & {\tiny 75\%} & {\tiny 74} & {\tiny 100\%} & {\tiny 33\%} & {\tiny 60} & {\tiny 60\%} & {\tiny 100\%}\tabularnewline
\hline 
{\tiny 88} & {\tiny 100\%} & {\tiny 75\%} & {\tiny 84} & {\tiny 100\%} & {\tiny 33\%} & {\tiny 62} & {\tiny 60\%} & {\tiny 83\%}\tabularnewline
\hline 
{\tiny 32} & {\tiny 100\%} & {\tiny 67\%} & {\tiny 93} & {\tiny 100\%} & {\tiny 33\%} & {\tiny 77} & {\tiny 40\%} & {\tiny 60\%}\tabularnewline
\hline 
{\tiny 43} & {\tiny 100\%} & {\tiny 67\%} & {\tiny 90} & {\tiny 100\%} & {\tiny 17\%} & {\tiny 70} & {\tiny 33\%} & {\tiny 50\%}\tabularnewline
\hline 
{\tiny 45} & {\tiny 100\%} & {\tiny 67\%} & {\tiny 52} & {\tiny 83\%} & {\tiny 83\%} &  &  & \tabularnewline
\hline 
\end{tabular}

\caption{\label{tab:Users-judging-the-time-effort}Users judging the time effort }
 
\end{table}
Figure~\ref{fig:Users-judging-the-time-effort} illustrates a histogram
users versus correctly judged time effort. From this figure it is
also evident that users are far better at judging the time effort
for simple tasks. 35 users of 53 (66\%) have managed to correctly
judge the time effort of all simple tasks, but only 5 out 53 (9\%)
have managed to correctly judge the difficulty of all complex tasks.

Table~\ref{tab:Users-judging-the-query-effort} shows the results
for users judging the query effort for simple and complex tasks. 4
out of 53 users (8\%) were able to judge the time effort right for
simple and complex tasks. 22 out of 53 users (42\%) managed to judge
the query effort for simple tasks right and at the same time were
not totally correct for the complex tasks. Five users (9\%) were able
to correctly judge all complex tasks while wrongly judging a number
of simple tasks.

\begin{figure}
\includegraphics[bb=63bp 80bp 600bp 520bp,clip,width=1\columnwidth]{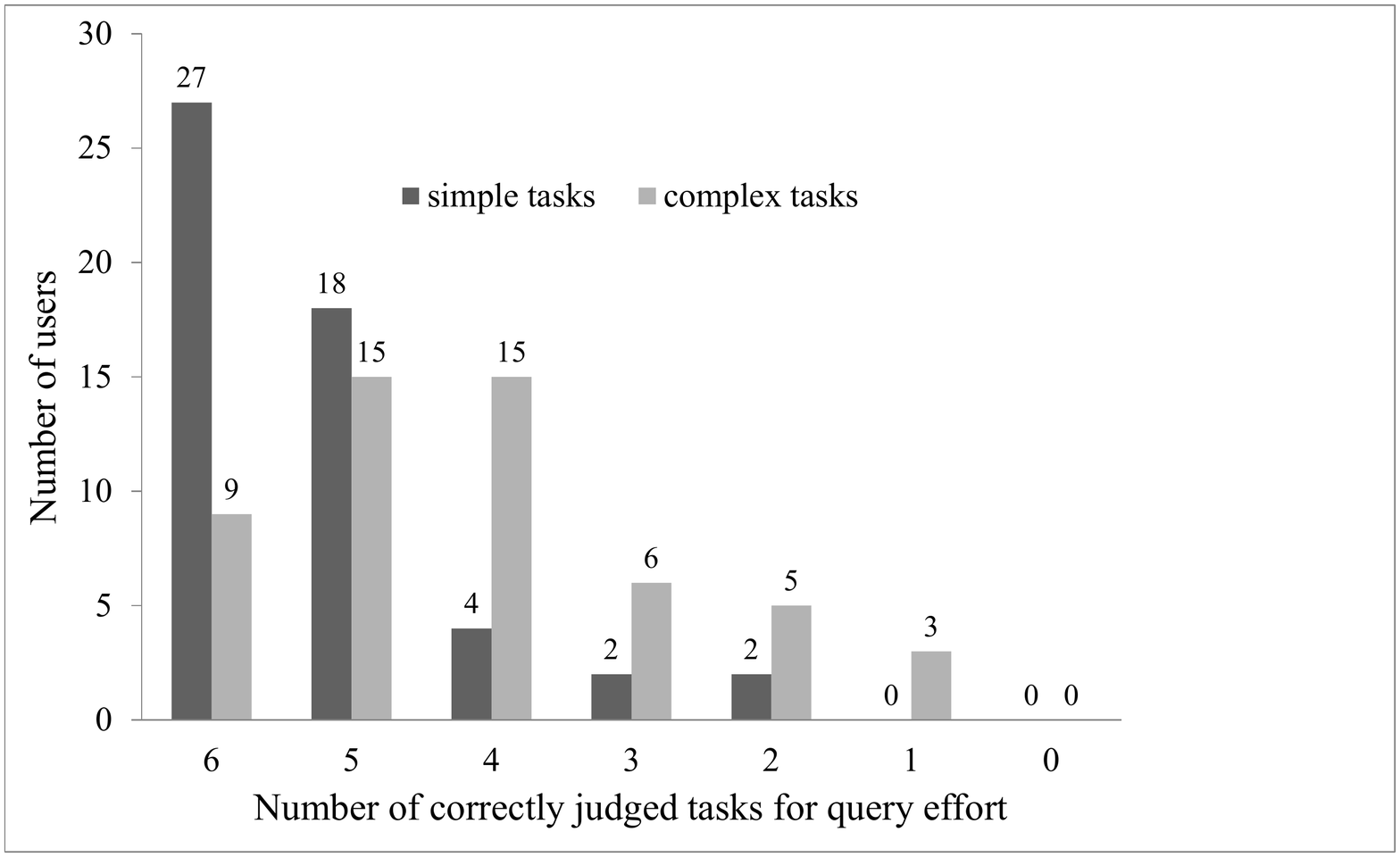}

\caption{\label{fig:Users-judging-the-query-effort}Users judging the query
effort}
\end{figure}

\begin{table}
\begin{tabular}{|>{\raggedright}p{0.03\columnwidth}|>{\raggedright}p{0.06\columnwidth}|>{\raggedright}p{0.08\columnwidth}||>{\raggedright}p{0.03\columnwidth}|>{\raggedright}p{0.06\columnwidth}|>{\raggedright}p{0.08\columnwidth}||>{\raggedright}p{0.03\columnwidth}|>{\raggedright}p{0.06\columnwidth}|>{\raggedright}p{0.08\columnwidth}|}
\hline 
{\tiny user} & {\tiny simple } & {\tiny complex} & {\tiny user} & {\tiny simple} & {\tiny complex} & {\tiny user} & {\tiny simple} & {\tiny complex}\tabularnewline
\hline 
\hline 
{\tiny 39} & {\tiny 100\%} & {\tiny 100\%} & {\tiny 78} & {\tiny 100\%} & {\tiny 50\%} & {\tiny 90} & {\tiny 83\%} & {\tiny 67\%}\tabularnewline
\hline 
{\tiny 88} & {\tiny 100\%} & {\tiny 100\%} & {\tiny 92} & {\tiny 100\%} & {\tiny 50\%} & {\tiny 79} & {\tiny 83\%} & {\tiny 67\%}\tabularnewline
\hline 
{\tiny 47} & {\tiny 100\%} & {\tiny 100\%} & {\tiny 61} & {\tiny 100\%} & {\tiny 50\%} & {\tiny 41} & {\tiny 83\%} & {\tiny 60\%}\tabularnewline
\hline 
{\tiny 69} & {\tiny 100\%} & {\tiny 100\%} & {\tiny 59} & {\tiny 100\%} & {\tiny 50\%} & {\tiny 44} & {\tiny 83\%} & {\tiny 50\%}\tabularnewline
\hline 
{\tiny 93} & {\tiny 100\%} & {\tiny 83\%} & {\tiny 24} & {\tiny 100\%} & {\tiny 40\%} & {\tiny 48} & {\tiny 83\%} & {\tiny 50\%}\tabularnewline
\hline 
{\tiny 42} & {\tiny 100\%} & {\tiny 83\%} & {\tiny 84} & {\tiny 100\%} & {\tiny 33\%} & {\tiny 75} & {\tiny 83\%} & {\tiny 17\%}\tabularnewline
\hline 
{\tiny 65} & {\tiny 100\%} & {\tiny 83\%} & {\tiny 82} & {\tiny 100\%} & {\tiny 33\%} & {\tiny 62} & {\tiny 80\%} & {\tiny 83\%}\tabularnewline
\hline 
{\tiny 63} & {\tiny 100\%} & {\tiny 83\%} & {\tiny 74} & {\tiny 100\%} & {\tiny 17\%} & {\tiny 46} & {\tiny 80\%} & {\tiny 80\%}\tabularnewline
\hline 
{\tiny 34} & {\tiny 100\%} & {\tiny 67\%} & {\tiny 43} & {\tiny 97\%} & {\tiny 67\%} & {\tiny 55} & {\tiny 80\%} & {\tiny 75\%}\tabularnewline
\hline 
{\tiny 89} & {\tiny 100\%} & {\tiny 67\%} & {\tiny 54} & {\tiny 88\%} & {\tiny 40\%} & {\tiny 68} & {\tiny 67\%} & {\tiny 83\%}\tabularnewline
\hline 
{\tiny 51} & {\tiny 100\%} & {\tiny 67\%} & {\tiny 28} & {\tiny 83\%} & {\tiny 100\%} & {\tiny 76} & {\tiny 67\%} & {\tiny 83\%}\tabularnewline
\hline 
{\tiny 87} & {\tiny 100\%} & {\tiny 67\%} & {\tiny 83} & {\tiny 83\%} & {\tiny 100\%} & {\tiny 67} & {\tiny 67\%} & {\tiny 80\%}\tabularnewline
\hline 
{\tiny 45} & {\tiny 100\%} & {\tiny 67\%} & {\tiny 72} & {\tiny 83\%} & {\tiny 100\%} & {\tiny 38} & {\tiny 60\%} & {\tiny 40\%}\tabularnewline
\hline 
{\tiny 73} & {\tiny 100\%} & {\tiny 67\%} & {\tiny 52} & {\tiny 83\%} & {\tiny 83\%} & {\tiny 70} & {\tiny 50\%} & {\tiny 100\%}\tabularnewline
\hline 
{\tiny 32} & {\tiny 100\%} & {\tiny 67\%} & {\tiny 37} & {\tiny 83\%} & {\tiny 83\%} & {\tiny 58} & {\tiny 50\%} & {\tiny 20\%}\tabularnewline
\hline 
{\tiny 64} & {\tiny 100\%} & {\tiny 60\%} & {\tiny 57} & {\tiny 83\%} & {\tiny 83\%} & {\tiny 60} & {\tiny 40\%} & {\tiny 100\%}\tabularnewline
\hline 
{\tiny 66} & {\tiny 100\%} & {\tiny 60\%} & {\tiny 71} & {\tiny 83\%} & {\tiny 75\%} & {\tiny 77} & {\tiny 40\%} & {\tiny 60\%}\tabularnewline
\hline 
{\tiny 33} & {\tiny 100\%} & {\tiny 60\%} & {\tiny 81} & {\tiny 83\%} & {\tiny 75\%} &  &  & \tabularnewline
\hline 
\end{tabular}

\caption{\label{tab:Users-judging-the-query-effort}Users judging the query
effort }
 
\end{table}
Figure~\ref{fig:Users-judging-the-query-effort} illustrates a histogram
users versus correctly judged query effort. From this figure it is
also evident that users are far better at judging the query effort
for simple tasks. 27 users of 53 (51\%) have managed to correctly
judge the query effort of all simple tasks, but only 9 out 53 (17\%)
have managed to correctly judge the difficulty of all complex tasks.

Table~\ref{tab:Users-judging-the-task-outcome} shows the results
for users judging the task outcome for simple and complex tasks. 14
out of 53 users (26\%) were able to judge the time effort right for
simple and complex tasks at the same time. 27 out of 53 users (51\%)
managed to judge the task outcome for simple tasks right and at the
same time were not totally correct for the complex tasks. One user
was able to correctly judge all complex tasks while wrongly judging
a number of simple tasks.

\begin{table}
\begin{tabular}{|>{\raggedright}p{0.03\columnwidth}|>{\raggedright}p{0.06\columnwidth}|>{\raggedright}p{0.08\columnwidth}||>{\raggedright}p{0.03\columnwidth}|>{\raggedright}p{0.06\columnwidth}|>{\raggedright}p{0.08\columnwidth}||>{\raggedright}p{0.03\columnwidth}|>{\raggedright}p{0.06\columnwidth}|>{\raggedright}p{0.08\columnwidth}|}
\hline 
{\tiny user} & {\tiny simple } & {\tiny complex} & {\tiny user} & {\tiny simple} & {\tiny complex} & {\tiny user} & {\tiny simple} & {\tiny complex}\tabularnewline
\hline 
\hline 
{\tiny 24} & {\tiny 100\%} & {\tiny 100\%} & {\tiny 73} & {\tiny 100\%} & {\tiny 83\%} & {\tiny 82} & {\tiny 100\%} & {\tiny 50\%}\tabularnewline
\hline 
{\tiny 32} & {\tiny 100\%} & {\tiny 100\%} & {\tiny 90} & {\tiny 100\%} & {\tiny 83\%} & {\tiny 87} & {\tiny 100\%} & {\tiny 50\%}\tabularnewline
\hline 
{\tiny 33} & {\tiny 100\%} & {\tiny 100\%} & {\tiny 48} & {\tiny 100\%} & {\tiny 80\%} & {\tiny 93} & {\tiny 100\%} & {\tiny 40\%}\tabularnewline
\hline 
{\tiny 34} & {\tiny 100\%} & {\tiny 100\%} & {\tiny 54} & {\tiny 100\%} & {\tiny 80\%} & {\tiny 46} & {\tiny 100\%} & {\tiny 33\%}\tabularnewline
\hline 
{\tiny 39} & {\tiny 100\%} & {\tiny 100\%} & {\tiny 74} & {\tiny 100\%} & {\tiny 80\%} & {\tiny 63} & {\tiny 100\%} & {\tiny 33\%}\tabularnewline
\hline 
{\tiny 43} & {\tiny 100\%} & {\tiny 100\%} & {\tiny 76} & {\tiny 100\%} & {\tiny 78\%} & {\tiny 60} & {\tiny 83\%} & {\tiny 83\%}\tabularnewline
\hline 
{\tiny 44} & {\tiny 100\%} & {\tiny 100\%} & {\tiny 55} & {\tiny 100\%} & {\tiny 75\%} & {\tiny 89} & {\tiny 83\%} & {\tiny 83\%}\tabularnewline
\hline 
{\tiny 52} & {\tiny 100\%} & {\tiny 100\%} & {\tiny 58} & {\tiny 100\%} & {\tiny 75\%} & {\tiny 64} & {\tiny 83\%} & {\tiny 67\%}\tabularnewline
\hline 
{\tiny 57} & {\tiny 100\%} & {\tiny 100\%} & {\tiny 28} & {\tiny 100\%} & {\tiny 67\%} & {\tiny 70} & {\tiny 83\%} & {\tiny 67\%}\tabularnewline
\hline 
{\tiny 71} & {\tiny 100\%} & {\tiny 100\%} & {\tiny 37} & {\tiny 100\%} & {\tiny 67\%} & {\tiny 75} & {\tiny 83\%} & {\tiny 63\%}\tabularnewline
\hline 
{\tiny 72} & {\tiny 100\%} & {\tiny 100\%} & {\tiny 38} & {\tiny 100\%} & {\tiny 67\%} & {\tiny 41} & {\tiny 83\%} & {\tiny 50\%}\tabularnewline
\hline 
{\tiny 77} & {\tiny 100\%} & {\tiny 100\%} & {\tiny 45} & {\tiny 100\%} & {\tiny 67\%} & {\tiny 47} & {\tiny 83\%} & {\tiny 33\%}\tabularnewline
\hline 
{\tiny 84} & {\tiny 100\%} & {\tiny 100\%} & {\tiny 51} & {\tiny 100\%} & {\tiny 67\%} & {\tiny 83} & {\tiny 83\%} & {\tiny 33\%}\tabularnewline
\hline 
{\tiny 92} & {\tiny 100\%} & {\tiny 100\%} & {\tiny 88} & {\tiny 100\%} & {\tiny 67\%} & {\tiny 81} & {\tiny 80\%} & {\tiny 67\%}\tabularnewline
\hline 
{\tiny 42} & {\tiny 100\%} & {\tiny 83\%} & {\tiny 61} & {\tiny 100\%} & {\tiny 50\%} & {\tiny 59} & {\tiny 80\%} & {\tiny 20\%}\tabularnewline
\hline 
{\tiny 62} & {\tiny 100\%} & {\tiny 83\%} & {\tiny 66} & {\tiny 100\%} & {\tiny 50\%} & {\tiny 79} & {\tiny 50\%} & {\tiny 67\%}\tabularnewline
\hline 
{\tiny 65} & {\tiny 100\%} & {\tiny 83\%} & {\tiny 68} & {\tiny 100\%} & {\tiny 50\%} & {\tiny 69} & {\tiny 40\%} & {\tiny 100\%}\tabularnewline
\hline 
{\tiny 67} & {\tiny 100\%} & {\tiny 83\%} & {\tiny 78} & {\tiny 100\%} & {\tiny 50\%} &  &  & \tabularnewline
\hline 
\end{tabular}

\caption{\label{tab:Users-judging-the-task-outcome}Users judging the task
outcome }
 
\end{table}

\begin{figure}
\includegraphics[bb=67bp 68bp 610bp 535bp,clip,width=1\columnwidth]{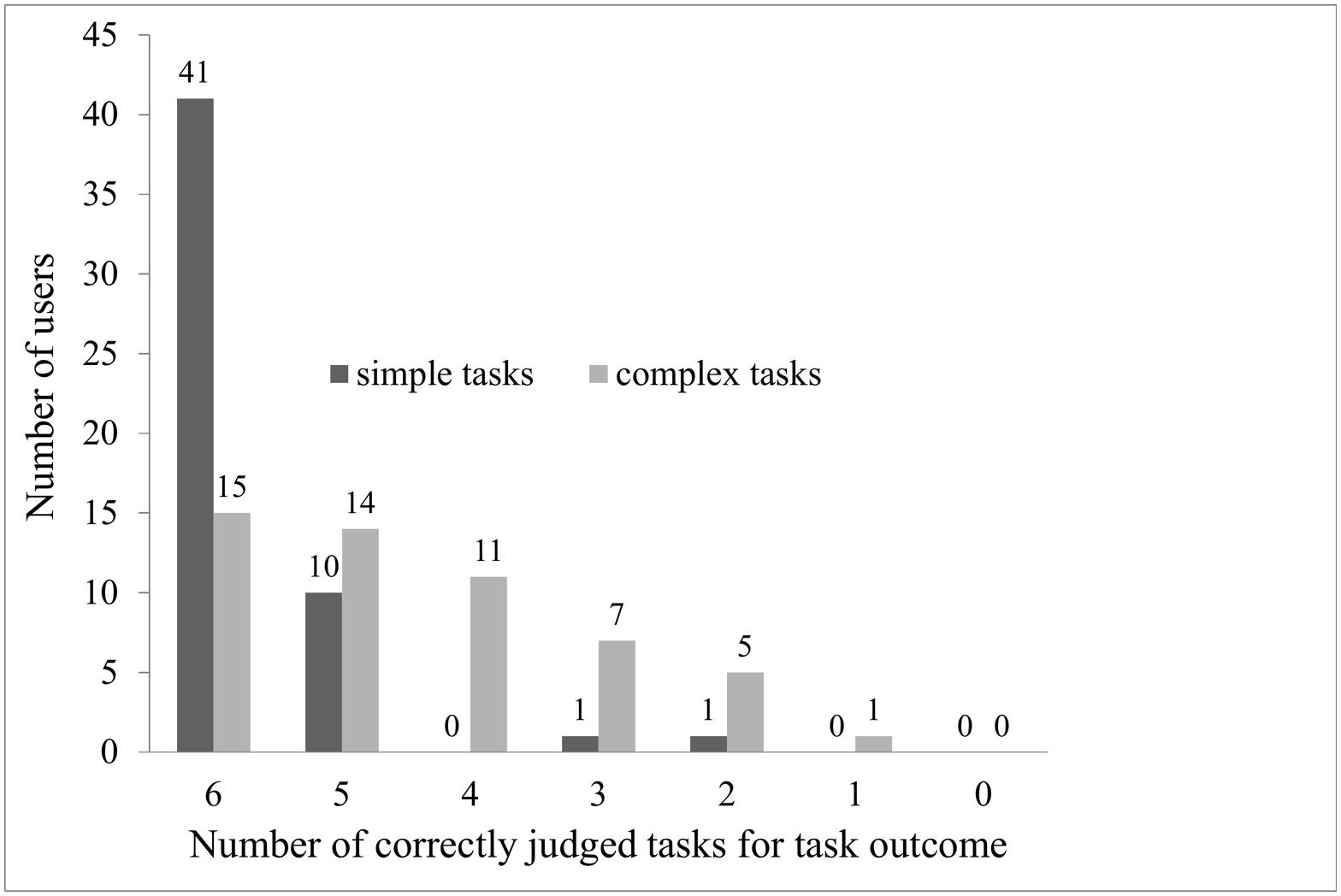}

\caption{\label{fig:Users-judging-the-task-outcome}Users judging the task
outcome}
\end{figure}

Figure~\ref{fig:Users-judging-the-task-outcome} illustrates a histogram
users versus correctly judged task outcome. From this figure it is
also evident that users are far better at judging the query effort
for simple tasks. 41 users of 53 (77\%) have managed to correctly
judge the task outcome of all simple tasks, but only 15 out 53 (28\%)
have managed to correctly judge the difficulty of all complex tasks.

\selectlanguage{american}%

\selectlanguage{english}%

\section{Discussion}

In this section, we first discuss our findings in the context of our
six research questions. Then, we discuss the performance of the users
with some individual search tasks that brought some interesting results
in terms of judgment capability.

It is obvious that in the case of simple tasks people are very well
capable of estimating, how difficult a simple search task would be.
For 90\% of the study participants the estimated and experienced difficulties
were in line. 

This might be due to the fact, that most people, even ordinary Web
users, have sufficient experience with carrying out simple search
tasks on the Internet. Therefore they know what to expect. This could
also be interpreted in a way that users also know, what search engines
can help them with, as far as simple tasks are concerned.  When
it comes to judging the search outcome and whether users would be
able to find the correct results, 95\% of the study participants correctly
assess their ability to find the correct result. Ability here needs
to be understood as comprising: as well understanding the problem
as carrying out the task with a search tool.

When examining users' ability to judge the aforementioned parameters
for complex search tasks, as expected their ability to judge ability
goes down in comparison to simple tasks. However, that close to 65\%
are still able to sufficiently judge the subjective difficulty was
a bit surprising to observe. It needs to be kept in mind that the
experiment was carried out in a laboratory environment and probably
the participants would judge differently in a real life scenario,
where they were more emotionally involved in the study. In addition,
especially the high factor of 73\% claiming to have found the correct
results is not in line with our manual evaluation of their results.
Only 47\% (158 out of 336 carried out tasks) of the results that were
submitted for complex search tasks were correct. This may indicate
that the problem with complex web searching might not be users finding
no results, but the results found only seemingly being correct. This
may explain why users are generally satisfied with their web search
outcomes.

As expected we have observed significant differences between users
judging simple tasks and users judging complex tasks. Users are significantly
better at judging simple tasks than at judging complex tasks.

When examining users' ability to judge whether they had carried out
a complex search task correctly or not, as in the previous section,
the difference is significant. It is interesting to observe that in
case of simple tasks, the users' judgment ability regarding the correctness
of the task outcome is over 3 times better than in case of complex
tasks (10\% error rate versus 33\% error rate). This could either
be due to the fact that the complex tasks were quite open (not specific
enough) as described by White and Iivonen \cite{white_assessing_2002}
and therefore the users did not really have a sense for correct and
incorrect. The other explanation would be that users got less support
from the search engine side than expected.

When it comes to search capabilities, we would expect that better
searchers would also be better at judging the difficulty, the effort
and the task outcome for complex search tasks. As the results shows,
only for the task outcome, users who perform better in the whole experiment
are also significantly better at judging the outcome of the task.
For difficulty and effort, the differences are insignificant. 

Regarding the question whether the judging performance is independent
of the task type (simple/complex) but depends on the user the answer
is as follows. There are some users who are able to correctly judge
the task parameters like task outcome (26\% of all users) both for
simple and for complex tasks. Yet the number of users who managed
to correctly judge those parameters for simple tasks (and were wrong
for all complex tasks) is much bigger (51\% in case of task outcome)
than the number of users who correctly judged the parameters for complex
tasks and at the same time were wrong with their judgments for simple
tasks (only one user in case of task outcome). Although numbers vary,
this relationship also holds true for task complexity, time effort,
and query effort. Together with the results from research questions
RQ1 to RQ5 it seems that task complexity indeed impacts the judging
performance of users.\selectlanguage{american}%

\selectlanguage{english}%

\section{Conclusions and Future Work\label{section:conclusions}}

In this paper we have examined how ordinary Web search engine users
manage to judge the three parameters task difficulty, task effort
and task outcome for search tasks. We have compared according judgments
for simple tasks and for complex tasks and also investigated, whether
better searchers would also be better judges. In addition we have
investigated, whether the judging performance depends on task complexity
or simply on the individual searcher.

Our results confirm that people are very well able to judge difficulty,
effort and task outcome for simple tasks. They are significantly less
good when they are asked to do the same for complex search tasks.
Users tend to over estimate their own search capabilities in case
of complex search tasks. We also observed, that better searchers are
also better at judging whether they would be able to find the correct
information than worse performing searchers. Regarding the hypothesis
that the judging performance might be depending on the users themselves
(and not the task type), we can conclude from our results that task
complexity is the main impacting factor for judging performance.

We also analyzed, how search engine operators could use the results
of this paper to offer better support for search engine users to assess
tasks. As expected in case of simple search tasks users are quite
well able to judge task parameters like difficulty and effort. We
have identified little need to offer better support for this kind
of tasks. Yet when it comes to complex search tasks, we think it would
help that search engines would at least build awareness about task
complexity. Users need to know, that tasks are different and that
their expectations need to be in line with task complexity. If a task
is complex, a user has to know, that he needs to put in more cognitive
effort as stated by Gwizdka \cite{gwizdka_assessing_2010}. We assume
that more awareness would lower dissatisfaction. In addition it is
also thinkable that search engine operators identify when people e.g.
work on a task over a longer time. They could then help those users
by offering estimates for task effort and task time based on similar
tasks carried out by other searchers. 

As far as limitations of our study are concerned, we think that the
sample used in our study was a bit wide. Users with very different
backgrounds (from the house wife to the university student) participated.
While this was of course intended to get realistic outcomes, it has
also resulted in sometimes high standard errors of mean for certain
indicators. Some of the studies mentioned in the related work section
(that only work with e.g. university students) might have brought
clearer results. Yet a wide validity of results of those studies (that
worked with less representative user samples) towards drawing conclusions
for mainstream users remains questionable.Another limitation is that
the study was carried out in laboratory environment and people were
only given a limited amount of time to carry out the tasks. We assume
that taking away the time limitation would lead to slightly different
results as published by Singer et al. \cite{singer_search-logger_2011}.

In future work it would be interesting to not only analyze if study
participants correctly or incorrectly judged tasks but also investigate
to what extent the users tend to over- and underestimate the task
parameters. Regarding sample size we are planning to run experiments
with bigger sample sizes. This will enable us to get more correct
statistics with more significant features. In addition we are planning
to conduct studies with study participants from certain professional
domains like teachers or blue collar workers only.

\subsection*{Acknowledgments}

This paper was supported by the European Union Regional Development
Fund through the Estonian Centre of Excellence in Computer Science
and by the target funding scheme SF0180008s12. In addition the research
was supported by the Estonian Information Technology Foundation (EITSA)
and the Tiger University program, as well as by Archimedes.\selectlanguage{american}%

\bibliographystyle{acm}
\bibliography{GeorgPhD}

\end{document}